# Evaluating and Improving ChatGPT-Based Expansion of Abbreviations

Yanjie Jiang, Hui Liu, Lu Zhang

*Abstract*—Source code identifiers often contain abbreviations. Such abbreviations may reduce the readability of the source code, which in turn hinders the maintenance of the software applications. To this end, accurate and automated approaches to expanding abbreviations in source code are desirable and abbreviation expansion has been intensively investigated. However, to the best of our knowledge, most existing approaches are heuristics, and none of them has even employed deep learning techniques, let alone the most advanced large language models (LLMs). LLMs have demonstrated cutting-edge performance in various software engineering tasks, and thus it has the potential to expand abbreviation automatically. To this end, in this paper, we present the first empirical study on LLM-based abbreviation expansion. Our evaluation results on a public benchmark suggest that ChatGPT is substantially less accurate than the state-of-the-art approach, reducing precision and recall by 28.2% and 27.8%, respectively. We manually analyzed the failed cases, and discovered the root causes for the failures: 1) Lack of contexts and 2) Inability to recognize abbreviations. In response to the first cause, we investigated the effect of various contexts and found surrounding source code is the best selection. In response to the second cause, we designed an iterative approach that identifies and explicitly marks missed abbreviations in prompts. Finally, we proposed a post-condition checking to exclude incorrect expansions that violate commonsense. All such measures together make ChatGPT-based abbreviation expansion comparable to the state of the art while avoiding expensive source code parsing and deep analysis that are indispensable for state-of-the-art approaches.

## I. INTRODUCTION

Identifiers play a crucial role in source code [27], frequently used to name various software entities such as classes, methods, and variables. Empirical studies [14] indicate that identifiers account for approximately 70 percent of source code characters. Given their substantial presence, the readability of these identifiers significantly impacts software quality, particularly in terms of maintainability and comprehension.

Abbreviations are common in identifiers [28], serving as concise representations of long terms or phrases. To shorten identifiers, developers often employ abbreviations (e.g., ww) instead of full terms (e.g., window wrapper) to name software entities, which results in a large number of abbreviations. However, the use of abbreviations also poses challenges, especially for newcomers to a codebase or those unfamiliar with specific domain terminology. If abbreviations are used improperly, they may find that it is often difficult to understand and interpret the abbreviations. The misinterpretation, in turn, may result in errors in software systems, reducing the quality of software systems.

To minimize the negative impact of abbreviations, several approaches have been proposed for expanding abbreviations in source code. An intuitive and straightforward approach is to look up words in dictionaries, such as generic English dictionaries. Such a dictionary-based approach is simple and straightforward, making it widely used. However, multiple dictionary words can match a single abbreviation, making it difficult to select the correct expansion. To improve accuracy, researchers have exploited the contexts in which the abbreviation appears, such as words within the same document or project, and names of software entities associated with the entity where the abbreviation appears. Such contextual approaches have significantly improved the performance of abbreviation expansion. However, such context-aware approaches heavily depend on contexts that are often resource-consuming to retrieve. For example, tfExpander [21] and kgExpander [22] depend on knowledge graphs of identifiers, and constructing such knowledge graphs requests parsing and deep analysis of the whole project. As a result, the construction may take a few minutes, which makes the context-aware approaches less useful in the industry because developers are loath to wait. Besides, they often request the source code be free from syntax errors because they may prevent the source code's parsing and accurate analysis.

With the advancements in machine learning and deep learning, particularly in the development of large language models (LLMs), the enhanced capabilities of machine learning techniques have facilitated breakthroughs in fields such as text summarization [24], translation [34], sentiment analysis [30], code search [32], code generation [17], and code summary [7]. Advanced LLMs, like ChatGTP, have proved excellent in understanding and generating both natural languages and source code [2], which makes them potentially capable of expanding abbreviations in source code. However, LLMs' ability to accurately expand abbreviations is not yet fully clear. Abbreviation expansion demands not only contextual awareness but also a deep understanding of semantic relationships among software entities, which presents a significant challenge.

To this end, in this paper, we present the first empirical study on LLM-based abbreviation expansion to investigate whether ChatGPT (one of the most advanced LLMs) is comparable to the specially designed algorithms in expanding abbreviations in source code. We incorporate ChatGPT to expand abbreviations in identifiers and analyze the quality of the generated new identifier to answer the following research questions.

*RQ1: Is ChatGPT accurate in expanding abbreviations in source code?* To answer this research question, we followed the widely used few-shot prompting and evaluated it on a public dataset. *Answer to RQ1:* Our evaluation results suggest that ChatGPT was substantially less accurate than the state-of-the-art approach although it correctly expanded more than half of the abbreviations.

*RQ2: Where and why does ChatGPT fail to expand abbreviations correctly?* To further improve the performance of ChatGPT-based abbreviation expansion, we manually checked each failed case, analyzed the root cause of ChatGPT's failure, and figured out how to address the limitation. *Answer to RQ2:* Our analysis suggests that most of the failures are due to 1) lacking of context and 2) failing to recognize abbreviations. Lacking contexts often resulted in full terms that are significantly different from the expected ones. Failing to recognize abbreviations led to low recall.

*RQ3: To what extent can we improve the performance of ChatGPT-based abbreviation expansion by including the contexts of the abbreviations in prompts?* Inspired by the answers to RQ2, we tried to improve ChatGPT-based abbreviation expansion by providing various contexts.

- *RQ3-1: To what extent can we improve the performance of ChatGPT-based abbreviation expansion by including the enclosing file of the abbreviations?* The first context we tried is the enclosing document of the abbreviation, containing numerous words, identifiers, and sentences that are potentially related to the abbreviation. *Answer to RQ3-1:* Including the enclosing file in the prompt can substantially improve the performance, improving the precision and recall by 19 and 20 percentage points, respectively. However, its precision (83%) and recall (81%) remain substantially lower than those of tfExpander (precision = 92% and recall = 89%).

- *RQ3-2: To what extent can we improve the performance of ChatGPT-based abbreviation expansion by including the knowledge graphs associated with the enclosing identifiers?* Knowledge graphs of abbreviations have proved to be one of the most useful contexts for abbreviation expansion [21]. To this end, we appended the graphs to the prompts. *Answer to RQ3-2:* Knowledge graphs have an even greater positive effect on the performance than enclosing files, improving precision and recall to 89% and 85%, respectively.

- *RQ3-3: To what extent can we improve the performance of ChatGPT-based abbreviation expansion by including the surrounding code of the abbreviations?* We appended the surrounding code of the abbreviation (i.e., the preceding three lines of code and the following three lines of code as well as the line where the abbreviation appeared). *Answer to RQ3-3:* Such simple contexts (surrounding code) surprisingly outweighed the complex knowledge graphs in helping abbreviation expansion. The resulting recall (87%) is close to that (89%) of tfExpander.

*RQ4: To what extent can we improve the performance of ChatGPT-based abbreviation expansion by explicitly marking missed abbreviations?* Inspired by the answers to RQ2, we tried to improve the performance by explicitly marking missed abbreviations and requested ChatGPT to expand them again. *Answer to RQ4:* By discovering the missed abbreviations with existing algorithms, and explicitly requesting ChatGPT to expand the missed abbreviations, we can further improve the performance of ChatGPT-based abbreviation expansion, especially the recall (by 2 percentage points).

*RQ5: To what extent can we further improve ChatGPT-based abbreviation expansion by post-condition checking?* While conducting the analysis to answer RQ2, we recognized that some of the expansions are violating commonsense, i.e., the abbreviations should be subsequences of the suggested full terms. Consequently, we tried to avoid such mistakes by filtering out such expansions. *Answer to RQ5:* Heuristics-based post-condition checking can avoid some incorrect expansions, which improved the precision by 2 percentage points.

We also observe that all such useful measures (i.e., including surrounding code, identifying and explicitly marking missed abbreviations, and post-condition checking) made the ChatGPT-based abbreviation expansion as accurate as the state-of-the-art approach (tfExpander). **A significant advance of this approach** is that it does not request expensive static code analysis and it works even if the source code contains compilation errors. It makes ChatGPT-based abbreviation expansion attractive because the state-of-the-art approaches have to scan the whole software applications to build knowledge graphs before they can be applied.

In summary, this paper makes the following contributions:

- **The first empirical study** on LLM-based abbreviation expansion, revealing the potential and limitations of ChatGPT in expanding abbreviations in source code. The replication package is publicly available at [4], enabling verification and reproduction.
- **A series of measures** that substantially improve the performance of ChatGPT-based abbreviation expansion, making it as accurate as the state-of-the-art approaches.

The rest of the paper is structured as follows. Section II introduces related work. Section III presents the design of the empirical study whereas Section IV presents the results. Section V discusses threats and implications whereas Section VI presents the conclusions and future work.

## II. RELATED WORK

### A. Abbreviation Expansion

Abbreviation expansion is to convert abbreviations in identifiers into dictionary words. An intuitive approach is to utilize abbreviation dictionaries [5] that provide lists of well-known abbreviations and their corresponding full terms. This dictionary-based approach is often highly accurate. However, since these dictionaries are usually constructed manually, they tend to be significantly limited in size [11]. Consequently, approaches that rely entirely on such dictionaries can expand only a small number of abbreviations, leading to low recall.

Generic English dictionaries are also used for expanding abbreviations [15]. Approaches based on generic dictionaries compare a given abbreviation against each term in the dictionary and return those that match according to predefined rules. Because generic dictionaries are significantly larger than abbreviation-specific dictionaries, generic dictionaries-based approaches can accommodate a broader range of abbreviations. However, the challenge emerges when multiple terms match a given abbreviation, making it difficult to select the correct one.

More advanced techniques have been developed to identify full terms within the context of abbreviations. For example, Corazza et al. [13], Lawrie et al. [26], and Madani et al. [29] suggest looking for full terms from comments in the source code. However, such approaches are limited by the fact that developers rarely write comments. Lawrie et al. [25] suggest searching within enclosing methods, while Hill et al. [20] propose a more comprehensive strategy that sequentially examines enclosing methods, classes, and projects. Abdulrahman Alatawi et al. [9] take a different approach by employing a Bayesian unigram-based inference model to identify full terms directly from the source code. To the best of our knowledge, Abdulrahman Alatawi et al. [10] are the first to apply natural language models to abbreviation expansion, and their evaluations suggest that these models are highly accurate. Despite these advancements, the contexts exploited by these methods are often coarse-grained and neglect semantic relationships, which can significantly impact the effectiveness of abbreviation expansion.

In addition to the generic approaches in the preceding paragraphs, Jiang et al. [23] propose a parameter-specific method designed specifically for expanding abbreviations in parameters. By focusing on this specific subset of identifier abbreviations, their approach significantly improves both precision and recall. However, it is limited to parameters and cannot be extended to other types of identifiers. Therefore, they further proposed a generic and accurate approach to expand abbreviations in all identifiers by leveraging semantic relation and transfer expansion [21]. They build a knowledge graph to represent entity relationships and search the graph for full terms. In addition, semantic-based expansions identified in one context can be transferred to other relevant contexts.

### B. Large Language Models

Large language models (LLMs) are advanced models that have been pre-trained on extensive corpus [31], [35]. To effectively utilize the vast amount of unlabeled training data [6], [12], LLMs are typically pre-trained using self-supervised learning objectives [16], [33], [36]. Most LLMs are based on the Transformer architecture [35], which comprises an encoder for input representation and a decoder for output generation. Existing LLMs are generally classified into three categories: encoder-only models [16], decoder-only models [33], and encoder-decoder models [36].

To improve generalization and better align models with human intentions on unfamiliar tasks, recent developments have integrated techniques like instruction tuning and reinforcement learning. A notable example is ChatGPT, an advanced LLM

---

**Basic Prompt for ChatGPT**

You are a smart code maintainer. You will be asked questions related to abbreviation expansion. You can mimic answering them in the background 15 times and provide me with the most frequently appearing answer. Furthermore, please strictly adhere to the output format specified in the question. There is no need to explain your answer. I am going to give you a Java identifier. You should output a new identifier by expanding all abbreviations in the input identifier without any explanation. Please ignore the length of the new identifier and strictly follow the format given in the examples.

Examples:
Input: "VariableName" "textEvt"
Output: "textEvent"

Input: "MethodName" "getPurchaseURL"
Output: "GetPurchaseUniformResourceLocator"

Input: "VariableName" "overlinePosStr"
Output: "overlinePositionString"

The given identifier is [identifier with abbreviations]

Fig. 1: Basic Prompt for Abbreviation Expansion

---

by OpenAI built on the GPT architecture. The model first undergoes instruction tuning and is then further refined using reinforcement learning from human feedback. Additionally, the emergence of open-source instruction-tuned LLMs has shown impressive performance across a diverse range of tasks, further highlighting the potential of these approaches.

## III. STUDY SETUP

### A. Benchmark

The empirical study is based on a public benchmark constructed by Jiang et al. [21]. The benchmark is composed of 2,253 abbreviations and their corresponding full terms. Notably, all items in the benchmark were constructed manually by human experts to ensure the quality [1]. The benchmark was selected because it was the largest publicly available dataset of abbreviations. Reusing a public benchmark instead of constructing a new one is beneficial. On one side, it may reduce the cost of the empirical study. On the other side, it may help reduce potential bias.

### B. Large Language Model

The empirical study selected ChatGTP-4o as the targeted large language model. ChatGPT-4o was selected because it represented the state of the art [3] and it has been widely employed for various software engineering tasks [8]. To automate the invocation of ChatGPT, we accessed it through OpenAI's API instead of the web-based query.

Notably, we did not fine-tune ChatGPT because it is unlikely that software developers would like to fine-tune a large language model especially for abbreviation expansion. Keeping a unique large language model, if possible, for all software engineering tasks could be much more efficient and convenient. To this end, we followed the widely used few-shot pattern to provide examples to ChatGTP so that it could learn instantly from the examples.

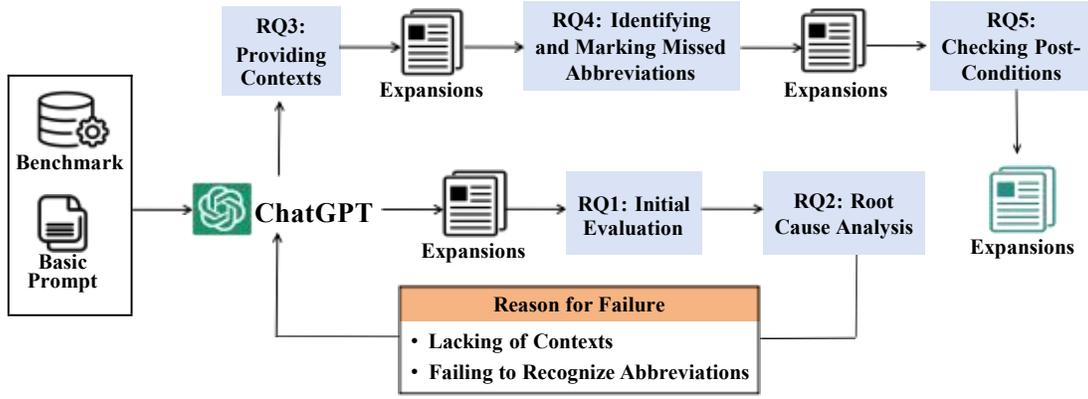

Fig. 2: Workflow of the Empirical Study

*C. Metrics*

To assess the performance in abbreviation expansion, we reused the performance metrics widely employed by related works [22], [23], including *precision*, *recall*, and $F_1$:

$$Precision = \frac{TP}{TP + FP} \quad (1)$$

$$Recall = \frac{TP}{TN} \quad (2)$$

$$F_1 = \frac{2 \times Precision \times Recall}{Precision + Recall} \quad (3)$$

where TP represents the true positive (correct expansion), FP represents the false positive (incorrect expansion) and TN represents the total number of abbreviations.

*D. Design of Basic Prompts*

To strike a balance between using overly simplistic prompts that might underestimate ChatGPT's capabilities and overly complex prompts that are uncommon in practical use, we have designed our basic prompt by carefully adhering to common practices and widely accepted experiences in the field of using ChatGPT. In particular, our basic prompt includes three parts:
- The natural language description part that explains the task.
- Examples.
- The identifier that contains abbreviations.

Figure 1 presents our basic prompt. Based on the widely-knowledged experience of using ChatGPT, the first part includes the following contents:
- A role-playing instruction to inspire ChatGPT's capability of abbreviation expansion. Such role-playing instruction is a common prompt optimization strategy [18].
- A task-description instruction to provide clear guidance and improve the accuracy and relevance of responses.

Notably, to reduce the randomness of ChatGPT predictions, we took the following measures. First, we requested ChatGTP to repeat the prediction multiple times by specifying in the prompt "*You can mimic answering them in the background 15 times and provide me with the most frequently appearing answer.*" Second, we repeated the evaluation five times and computed the average performance as the final performance to be presented in the paper.

*E. Experimental Procedure*

Figure 2 presents the overview of our experimental procedure. In the first place, we evaluated ChatGPT-based abbreviation expansion with the basic prompt designed in Section III-D. We computed its performance metrics, i.e., precision, recall, and $F_1$ score. Such metrics may answer research question RQ1. After that, we manually analyzed the cases where ChatGPT failed with the default prompt, aiming to identify the root causes for the failures. As a result of the analysis, we answered research question RQ2 by providing a list of *reasons for failure*, i.e., lacking contexts and failing to recognize abbreviations.

Based on the analysis results of RQ2, we took a series of measures to resolve the problems. First, to resolve the problem of "*lacking contexts*", we included various contexts to the prompt, including enclosing files (RQ3.1), knowledge graphs (RQ3.2), and surrounding code lines (RQ3.3). The evaluation results suggest that surrounding code lines have the greatest potential, and thus in the following evaluation, we took it as the baseline and tried to further improve it with additional measures. Second, to resolve the problem of "*failing to recognize abbreviations*", we identified remaining abbreviations in the output of ChatGPT with existing approaches, explicitly marked them in the prompt, and requested ChatGTP to expand them again, which may answer research question RQ4. Third, we tried to further improve ChatGPT-based abbreviation expansion by appending a post-condition checking, that removed all expansions that were not supersequences

TABLE I: Initial Evaluation of ChatGPT

|  | ChatGPT | tfExpander |
|---|---|---|
| # Abbreviations | 2,253 | 2,253 |
| # Expanded Abbreviations | 2,160 | 2,179 |
| # Correctly Expanded Abbreviations | 1,379 | 2,005 |
| # Incorrect Expansions | 781 | 174 |
| # Missed Abbreviations | 93 | 74 |
| Precision | 64% | 92% |
| Recall | 61% | 89% |
| $F_1$ | 62% | 90% |

```
1  public UpdateManager(BridgeContext ctx ...) {
2    ...
3    for (...) {
4      BridgeContext resCtx = secondaryBridgeContexts[ I ];
5      if (!(( SVGOMDocument) resCtx.getDocument())){
6        continue ;
7      }
8      resCtx . setUpdateManager(this) ;
9      ScriptingEnvironment se =
10              initializeScriptingEnvironment (resCtx) ;
11     secondaryScriptingEnvironments [ i ] = see;
12     ...
13   }
```

Listing 1: An Example of Outrageous Expansions

(containing sequences) of their according abbreviations. This attempt may answer research question RQ5.

IV. RESULTS AND ANLYSIS

A. RQ1: Initial Evaluation of ChatGPT

*1) Setup:* To answer RQ1, we automatically generated a unique prompt for each identifier in the benchmark. The prompt generation was based on the prompt template designed in Section III-D. We fed the generated prompt to ChatGPT-4o, and compared the output against the reference answers in the benchmark. An expansion is correct if and only if it is identical to the reference expansion in the benchmark.

*2) Results:* Evaluation results are presented in Table I. To contextualize the performance, we also present the result of tfExpander that represents the state of the art. From this table, we make the following observations:

- First, ChatGPT has great potential in expanding abbreviations in source code. It successfully expanded more than half of the abbreviations without any fine-tuning or adaption.
- Second, ChatGPT is by far less accurate than the state-of-the-art approach. Its precision (64%) and recall (61%) are substantially lower than those of tfExpander (precision =92% and recall=89%). It may suggest that even the most advanced generic large language models like ChatGPT cannot replace specialized tools in abbreviation expansion without essential adaption.
- The number of incorrect expansions (781) is substantially larger than that (174) of tfExpander, which leads to a substantial reduction in precision. To make the ChatGPT-based abbreviation expansion practical, we should find ways to correct such unexpected expansions.

In conclusion, ChatGPT has the potential for abbreviation expansion, but its performance remains far below the state of the art.

B. RQ2: Root Cause Analysis

*1) Setup:* To answer RQ2, we manually analyzed the 781 incorrect expansions as well as the 93 missed abbreviations (i.e., abbreviations that ChatGPT did not expand).

For the missed abbreviations, we explicitly marked them in the prompt and requested ChatGPT to expand them again. If the missed abbreviations were expanded successfully at this time, we say the root cause of the failure is ChatGPT's inability to recognize abbreviations. Otherwise, the root cause is ChatGPT's inability to figure out the correct full terms for the given abbreviations.

For each incorrect expansion, the analysis should answer the following questions:

- **Q1:** Is the suggested full term essentially different from the expected one? If they are different forms of the same word or are (near) synonyms of each other, we say that are not essentially different. Otherwise, they are.
- **Q2:** If the suggested full term is essentially different from the expected one, is it semantically related to the source code? The weaker the semantics relation is, the more absurd the mistake is.
- **Q3:** Can we find the expected full term within the contexts of the abbreviation, i.e., the enclosing document and its associated knowledge graph?

*2) Results:* ChatGPT's inability to recognize abbreviations is the major reason for missed abbreviations. Out of the 93 missed abbreviations, 62 were expanded after being explicitly marked in the prompts, with 49 of these expansions being correct. The results may suggest that 53%=49/93 of the missed abbreviations could be expanded correctly by simply marking the abbreviations in the prompts. If the accuracy of the LLM-based abbreviation expansion could be further improved (as suggested in the following sections), we can rescue even more missed abbreviations. Consequently, identifying and explicitly marking missed abbreviations in prompts is potentially a good way to improve LLM-based abbreviation expansion.

Analysis of the incorrectly expanded abbreviations suggests that 88%=688/781 of the incorrect expansions are essentially different from their reference expansions (answering question Q1). A typical example is presented in Listing 1 where ChatGPT expanded the abbreviation "*se*" on Line 9 into "*searchEngine*" that is essentially different from the reference expansion "*ScriptingEnvironment*". "*searchEngine*" and "*ScriptingEnvironment*" have completely different roots and etymologies, and they are not (near) synonyms of each other. In a conclusion, from the suggested expansion "*searchEngine*" we cannot correctly interpret the semantics of the variable

```
1  public long getWaitTime() {
2      long wt = waitTime;
3      waitTime = Long.MAX_VALUE;
4      return wt;
5  }
```

Listing 2: Expansion That Is Abrupt to Contexts

```
1  public static AffineTransform getViewTransform (...) {
2      ...
3      // 'viewBox'
4      float [] vb;
5      if (vh.hasViewBox) {
6          vb = vh.viewBox;
7      } else {
8          ...
9  }
```

Listing 3: Full Terms Appering in Contexts

TABLE II: Effect of Enclosing Files

|  | Default | + Enclosing File | Δ |
|---|---|---|---|
| # Abbreviations | 2,253 | 2,253 | 0 – |
| #Expanded Abb. | 2,160 | 2,177 | 17 ↑ |
| #Correct Expansions | 1,379 | 1,816 | 437 ↑ |
| #Incorrect Expansions | 781 | 361 | 420 ↓ |
| #Missed Abb. | 93 | 76 | 17 ↓ |
| Precision | 64% | 83% | 19 pp ↑ |
| Recall | 61% | 81% | 20 pp ↑ |
| $F_1$ | 62% | 82% | 20 pp ↑ |

"*se*", and thus the suggested expansion is misleading. For convenience, we call such suggested expansions that are essentially different from their corresponding reference expansions as *outrageous expansions*.

Answers to question Q2 suggest that 97%=665/688 *outrageous expansions* are not semantically related to the source code, especially its surrounding code. A typical example is presented in Listing 2. The abbreviation "*wt*" on Line 2 was expanded incorrectly into "*waterTemperature*". However, by looking into the source code, especially the surrounding code of the abbreviation, we may find that the source code does not mention anything that is related to *water temperature*. Consequently, if ChatGPT can learn from the surrounding code, it may have identified this problem, and thus may not suggest such an outrageous expansion. To validate the assumption, we appended the source code to the prompt and requested ChatGPT to expand again. With the updated prompt, ChatGPT correctly expanded the abbreviation into "*waitTime*" that is identical to the reference expansion. The analysis results may suggest that we could avoid (or reduce) outrageous expansions by appending surrounding code into the prompts.

For 617 out of the 688 *outrageous expansions*, we successfully found the expected full terms within the surrounding code lines of the abbreviation and/or the knowledge graphs [21] associated with the identifiers where the to-be-expanded abbreviations appear (answering question Q3). A typical example is presented in Listing 3 where the abbreviation "*vb*" on Line 4 was expanded by ChatGPT into "*variable*". At the first look, the expansion looks good since the abbreviation is used to name a *variable*. However, by looking into the whole document, we found that the variable *vb* is declared to store a view box by explicitly marking the coordinates and size of the view box. Consequently, the correct expansion should be "*viewBox*". we also noticed that such full terms appear multiple times in the surrounding code of the abbreviation, i.e., on Lines 3, 5, and 6. All such analyses may suggest that the contexts of abbreviations could be useful for ChatGPT-based abbreviation expansion.

Besides the preceding findings, we also observed that some of the suggested expansions violate well-known common-sense: The abbreviation should be a subsequence of its corresponding full term. If the expansions do not conform to this common knowledge, such expansions are deemed incorrect. In total, we found that 39 expansions suggested by ChatGPT violated this common sense. A typical example is "*dtde*" that was expanded by ChatGPT into "*dragDropEnd*". The expansion does not even contain the character 't' that appears in the original identifier.

we conclude based on the preceding analysis that lacking of contexts as well as the inability to recognize abbreviations are the major reasons for the failures of ChatGPT in abbreviation expansion.

### C. RQ3-1: Enclosing File as Contexts

*1) Setup:* As suggested by the analysis in Section IV-B, the contexts of the abbreviations could be potentially useful for abbreviation expansion. To this end, we appended to the prompt the whole file where the to-be-expanded abbreviation appears, called *enclosing file*. With this updated prompt, we repeated the evaluation of ChatGTP-based abbreviation expansion with exactly the same benchmark.

*2) Results:* The evaluation results are presented in Table II. The second column "*Default*" presents the results of ChatGPT without contexts (i.e., the results presented in Section IV-A). The third column "*+ Enclosing File*" presents the results when we appended the enclosing files to prompts. The last column "Δ" presents the changes caused by the addition of enclosing files, i.e., to what extent the enclosing files improve the performance.

From Table II, we observe that the performance of ChatGPT-based abbreviation expansion has been substantially improved when enclosing files were appended to prompts. It improved the precision from 64% to 83%, resulting in a substantial improvement of 19 percentage points. The same is true for recall: It was improved from 61% to 81% with an improvement of 20 percentage points. From the table, we also observe that the enclosing files have a positive effect on all performance metrics. It improved the number of expanded abbreviations and the number of correct expansions whereas the number of incorrect expansions as well as the number of missed abbreviations were reduced.

TABLE III: Effect of Knowledge Graphs

|  | Default | + Knowledge Graph | Δ |
|---|---|---|---|
| # Abbreviations | 2,253 | 2,253 | 0 – |
| # Expanded Abb. | 2,160 | 2,140 | 20 ↓ |
| # Correct Expansions | 1,379 | 1,915 | 536 ↑ |
| # Incorrect expansions | 781 | 225 | 556 ↓ |
| # Missed Abb. | 93 | 113 | 20 ↑ |
| Precision | 64% | 89% | 25 pp ↑ |
| Recall | 61% | 85% | 24 pp ↑ |
| $F_1$ | 62% | 87% | 25 pp ↑ |

TABLE IV: Effect of Surrounding Source Code

|  | Default | + Surrounding Source Code | Δ |
|---|---|---|---|
| # Abbreviations | 2,253 | 2,253 | 0 – |
| # Expanded Abb. | 2,160 | 2,190 | 30 ↑ |
| # Correct Expansions | 1,379 | 1,959 | 580 ↑ |
| # Incorrect Expansions | 781 | 231 | 550 ↓ |
| # Missed Abb. | 93 | 63 | 30 ↓ |
| Precision | 64% | 89% | 25 pp ↑ |
| Recall | 61% | 87% | 26 pp ↑ |
| $F_1$ | 62% | 88% | 26 pp ↑ |

Notably, including enclosing files in prompts has a greater influence on the number of incorrect expansions than the number of missed abbreviations. The former was reduced by 54%=420/781 whereas the latter was reduced by 18%=17/93 only. The results are reasonable. As suggested by the analysis results in Section IV-B, the incorrect expansions were generated by ChatGTP because of lacking contexts. Consequently, including the contexts (the enclosing files here) helps alleviate this problem, and thus could substantially reduce the number of incorrect expansions. In contrast, the contexts have little help in the identification of abbreviations, and thus appending the enclosing files to prompt failed to substantially increase the ability of ChatGPT in automated identification of abbreviations.

Including enclosing files in prompts may also result in negative effects although the metrics in Table II fail to reveal such negative effects. By checking and comparing the results in this section and those in Section IV-A, we found that including enclosing files in prompts made 53 previously successful expansions incorrect, although it rescued 490 previously failed expansions.

### D. RQ3-2: Knowledge Graph as Contexts

*1) Setup:* As suggested by Jiang et al. [22], knowledge graphs of the enclosing identifiers often contain full terms of the to-be-expanded abbreviations. Consequently, in this section, we tried to append such knowledge graphs to the prompts and requested ChatGPT to expand abbreviations with the updated prompts. Notably, the benchmark used in this case study was constructed by the same authors of the knowledge graphs (i.e., Jiang et al. [22]), and the benchmark is readily accompanied by the knowledge graphs. Consequently, we simply retrieved the knowledge graphs associated with the to-be-expanded abbreviations and appended them to the corresponding prompts.

*2) Results:* Table III presents the effect of including knowledge graphs in prompts for ChatGPT-based abbreviation expansion. From this table, we make the following observations:

- First, it substantially improved the performance. The improvement (upon the default setting) in recall, precision, and $F_1$ is 24, 25, and 25 percentage points, respectively. The relative improvement is 40% ($F_1$), 39% (recall), and 39% (precision), respectively. Such a great improvement may suggest that knowledge graphs are not only useful for traditional heuristics-based abbreviation expansion but also useful for LLM-based abbreviation expansion.
- The resulting improvement is even greater than that caused by adding enclosing files (as shown in Table II). One possible reason is that a knowledge graph represents the selected contexts of an abbreviation and such contexts have been manually selected and validated by human experts [21]. In contrast, the enclosing file of an abbreviation represents the raw data of its contexts, which is potentially useful but may also contain useless or redundant contents.
- Knowledge graphs are often much more compact than enclosing files. The average size of a knowledge graph is 20 tokens whereas the average size of an enclosing file is 4,222 tokens. Compact contexts are highly preferred if they are as effective as alternatives because 1) ChatGPT charges based on the size of the prompts, and thus compact contexts (resulting in short prompts) help reduce the cost of LLM-based abbreviation expansion and 2) LLMs like ChatGPT have been proved less accurate when prompts are lengthy [19], which forces LLMs to set strict limitations on prompts' size.

While knowledge graphs offer significant advantages as useful contexts for abbreviation expansion, they also suffer from serious drawbacks compared to enclosing files: It is often expensive to retrieve the knowledge graphs. As introduced by Jiang et al. [22], to construct the knowledge graph we have to scan the whole software application to identify all software entities that are semantically related to the entity whose name contains the to-be-expanded abbreviation. One on side, the static code analysis depends on source code parsing, and thus we may fail to retrieve the knowledge graph from incomplete or illegal source code. On the other side, static code analysis is resource-consuming, especially when the enclosing application is large. In contrast, the enclosing files are always ready.

### E. RQ3-3: Surrounding Code as Contexts

*1) Setup:* While knowledge graphs are expensive to construct and enclosing files are lengthy, we would like to select the most relevant part from the enclosing files as the contexts for LLM-based abbreviation expansion. As an initial try, we selected the source code surrounding the given abbreviation as the compact contexts, i.e., the preceding and following N lines of source code as well as the line where the abbreviation appears. So, the surrounding source code should be composed of N+1+N lines of source code. However, if the number of preceding (or following) code lines was less than N, we only took all of the preceding/following code lines, and thus the total size of the resulting surrounding code lines could be smaller than N+1+N.

*2) Results:* Table IV and Fig.3 present the effect of including surrounding source code (where N equals three) as contexts. From the table and the figure, we surprisingly find that the surrounding source code is even more effective than the complete enclosing files. The resulting precision (89%) and recall (87%) are substantially greater than those (precision =83% and recall= 81%) when the enclosing files are taken as contexts. The relative improvement in precision and recall is up to 7.2%=(89%-83%)/83% and 7.4%=(87%-81%)/81%, respectively. The results may suggest that replacing enclosing files with only surrounding source code as the contexts may result in substantial improvement in the performance of LLM-based abbreviation expansion. One possible reason for the improvement is that in most cases the (compact) surrounding code is sufficient to contextualize the abbreviation and thus the other (lengthy) parts of the enclosing file could be removed safely. Including such useless but lengthy parts in the prompt, unfortunately, prevents LLMs from identifying the most useful contexts from the lengthy prompt.

By comparing Table IV against Table III, we notice that the surrounding source code is as useful as (if not better than) the knowledge graphs. Both of them improved the precision by 25 percentage points whereas surrounding code resulted in slightly greater improvement in recall (26 pp vs 24 pp). A direct and visual comparison is available in Fig. 3. The results may suggest that we can replace the expensive knowledge graphs with easy-to-collect and easy-to-follow surrounding source code. The replacement can significantly increase the usability and efficiency of LLM-based abbreviation expansion.

To calibrate the setting of N that specifies the scope of the surrounding source code, we changed the value of N and repeated the evaluation. The evaluation results are presented in Fig. 4. From this figure, we observe that N = 3 results in the largest $F_1$. Consequently, in the rest of the paper, we take it as the default setting.

We conclude based on the preceding analysis that various contexts may help LLM-based abbreviation expansion. Among all such contexts that have been investigated by the study, surrounding code lines are the most effective.

### F. RQ4: Identifying and Marking Missed Abbreviations

*1) Setup:* As suggested by the manual analysis in Section IV-B, one of the major reasons for ChatGPT's failures is its inability to recognize abbreviations. To this end, to further improve ChatGPT in abbreviation expansion, we proposed a new approach that works as follows for each abbreviation abb:
- First-round expansion

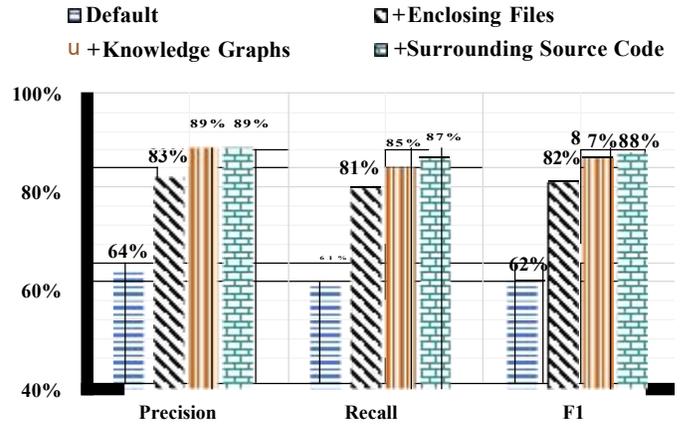

Fig. 3: Comparison Among Various Contexts

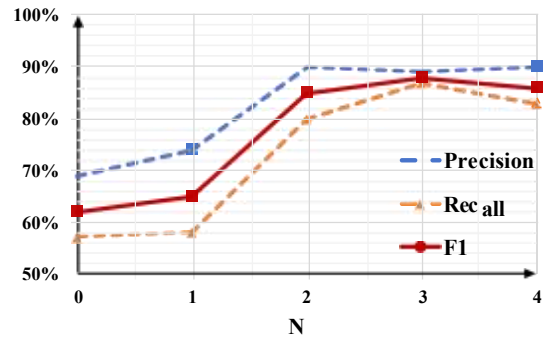

Fig. 4: Different Settings of N

  – Generate a unique prompt for abb according to the prompt template designed in Section III-D;
  – Retrieve its surrounding source code lines and append it to the resulting prompt;
  – Feed the updated prompt to ChatGPT and receive its output exp;
  – If exp does not contain any abbreviations, it terminates and exp is presented as the final output.
- Second-round expansion
  – Identify abbreviations in exp, and update the prompt by explicitly specifying the identified abbreviations and requesting ChatGPT to expand them;
  – Feed the updated prompt to ChatGPT and receive the output exp'. It is presented as the final expansion.

Notably, this approach is based on the *few-shot + surrounding source code* (i.e., the setting proposed in Section IV-E) instead of the initial baseline (i.e., *few-shot without any contexts*). For convenience, we call this new baseline as **Baseline #2** to distinguish it from the initial baseline proposed in Section IV-A. We also call this new approach (i.e., *few-shot + surrounding source code + marking missed abbreviations*) as **Baseline #3**.

*2) Results:* Table V illustrates how we can further improve Baseline #2 by identifying and marking missed abbreviations. The second column presents the performance of *Baseline #2* (i.e., *few-shot + surrounding source code*). The third column presents the performance of the new approach, i.e., *Baseline #3* that equals to *Baseline #2 + marking missed abbreviations*. The last column presents to what extent *Baseline #3* outperformed *Baseline #2*.

TABLE V: Effect of Marking Missed Abbreviations

|  | Baseline #2 | + Marking Missed Abb. | Δ |
|---|---|---|---|
| # Abbreviations | 2,253 | 2,253 | 0 – |
| # Expanded Abb. | 2,190 | 2,237 | 47 ↑ |
| # Correct Expan. | 1,959 | 2,003 | 44 ↑ |
| # Incorrect Expan. | 231 | 234 | 3 ↑ |
| # Missed Abb. | 63 | 16 | 47 ↓ |
| Precision | 89% | 90% | 1 pp ↑ |
| Recall | 87% | 89% | 2 pp ↑ |
| $F_1$ | 88% | 89% | 1 pp ↑ |

TABLE VI: Second-Round Expansion of Missed Abb.

| Items | Values |
|---|---|
| # Missed Abbreviations (first round) | 63 |
| # Identified Abbreviation | 63 |
| # Correct Expansions (second round) | 44 |
| # Incorrect Expansions (second round) | 3 |
| # Remaining Abbreviations (after second round) | 16 |

TABLE VII: Effect of Post-Condition Checking

|  | Baseline #3 | + Post-Condition Checking | Δ |
|---|---|---|---|
| # Abbreviations | 2,253 | 2,253 | 0 – |
| # Expanded Abb. | 2,237 | 2,182 | 55 ↓ |
| **# Violating Expan.** | 55 | 0 | 55 ↓ |
| # Correct Expan. | 2,003 | 2,003 | 0 – |
| # Incorrect Expan. | 234 | 179 | 55 ↓ |
| # Missed Abb. | 16 | 71 | 55 ↑ |
| Precision | 90% | 92% | 2 pp ↑ |
| Recall | 89% | 89% | 0 pp – |
| $F_1$ | 89% | 90% | 1 pp ↑ |

From Table V, we observe that we can slightly improve the performance of ChatGPT by identifying and marking missed abbreviations. Compared to *Baseline #2*, it improved the precision and recall by one and two percentage points, respectively. Although the improvement is minor, it did not cause any negative impact on any performance metrics on Table V, suggesting that it was both safe and effective.

To further investigate the reasons for the improvement, we retrieved all abbreviations missed by the first-round expansion (i.e., *Bseline #2*), and double-checked their expansion in the second-round expansion where the missed abbreviations were explicitly marked. The results of the analysis are presented in Table VI. From this table, we observe that the first-round expansion missed 63 abbreviations. The proposed approach successfully identified all such abbreviations by reusing the algorithm proposed by Jiang et al. [23] and explicitly marked them in their corresponding prompts. The second-round expansion (with the updated prompts) correctly expanded 70%=44/63 of the missed abbreviations, which improved the overall recall of the approach by two percentage points. A typical example is abbreviation "*Multi*" in method name "*testWithMultiDimensionalArray*". Explicitly marking the abbreviation forced ChatGPT to expand it correctly into "*Multiple*".

We also observed that 3 missed abbreviations were expanded incorrectly and 16 were missed again. The three incorrectly expanded abbreviations are "*tb*" (reference expansion "*toolBar*"), "*ACC*" (reference expansion "*Accumulator*"), and "*dfc*" (reference expansion "*DetectorFactoryCollection*"). ChatGPT expanded them into "*table*", "*account*", and "*DataFlowControl*", respectively. One possible reason for the 16 remaining abbreviations is the incapability of ChatGPT to expand them even though it knows exactly that they are abbreviations. Typical examples of such remaining abbreviations include "*fc*", "*pd*", and "*coll*". Such short abbreviations are often more challenging to expand than longer ones.

We conclude based on the preceding analysis that around three-quarters of the missed abbreviations could be expanded correctly if we can identify and mark such abbreviations explicitly in the prompts.

*G. RQ5: Post-Condition Checking*

*1) Setup:* As observed in Section IV-B, some of the suggested expansions violate common sense, i.e., an abbreviation should be a subsequence of its full term. Such expansions are deemed incorrect (although we may not exactly know their correct expansions), and thus excluding them may further improve the precision of ChatGPT in abbreviation expansion. To this end, we appended a post-condition checking step to *Baseline #3* (i.e., *few-shot + surrounding source code + marking missed abbreviations*) as follows:

- For each abbreviation to be expanded, we expand it with *Baseline #3*. The output is noted as exp;
- If none of the expansions in exp violates the post-condition, i.e., an abbreviation should be a subsequence of its full term, the expansion terminates and exp is returned as the final output;
- Expansions in exp that violate the post-condition are reversed, i.e., replaced with their corresponding abbreviations.
- The resulting exp is returned as the final expansion.

For convenience, we call this variant of *Baseline #3* as *Baseline #4*. Notably, *Baseline #4* does not make any additional invocation of ChatGPT. All that it does is to exclude some incorrect expansions by post-condition checking.

*2) Results:* Our evaluation results are presented in Table VII. The second and the third columns present the performance of *Baseline #3* and *Baseline #4* (i.e., *Baseline #3 + post-condition checking*), respectively. The last column presents how *post-condition checking* further improves *Baseline #3*.

From Table VII, we observe that the post-condition checking improved the precision by two percentage points without any reduction in recall. $F_1$ was improved slightly by one percentage point. Further analysis suggests that *Baseline #3* resulted in 55 *violating expansions* that failed to pass the post-condition checking. *Baseline #4* excluded all of them, and thus decreased the number of incorrect expansions by 55, which in turn increased the precision by two percentage points.

We noticed that the number of missed abbreviations was increased by 55. The reason is explained as follows. When we counted the number of missed abbreviations, we did not consider the correctness of their expansions. That is, if an abbreviation was expanded incorrectly, it would not be counted as a missed abbreviation. Consequently, when we removed the 55 incorrect expansions, their corresponding abbreviations became *missed abbreviations* because the approach did not provide any expansion for them.

We conclude based on the preceding analysis that the post-condition checking is convenient (with heuristics-based checking only) and effective.

## V. Discussions

### A. Threats To Validity

The first threat to external validity is that the case study was conducted on only 2,253 abbreviations. Special characters of the dataset could have biased the conclusions. To reduce the threat, we reused the largest benchmark that is publicly available for abbreviation expansion. We also notice that this benchmark has good diversity, covering identifiers from different projects and containing various abbreviations.

The second threat to external validity is that the case study is confined to Java identifiers. Conclusions drawn on Java are not necessarily true for other programming languages. The study is confined to Java because 1) the benchmark contains only Java identifiers and 2) the baseline approach (*tfExpander*) handles only Java identifiers.

The third threat to external validity is that only a single large language model is involved in the study. Conclusions drawn on this model may not hold for other large language models. To minimize the threat, we selected ChatGPT-4o, the latest version of the most advanced large language model. Notably, the case study was time and resource-consuming, and thus involving additional large language models is difficult and expensive, which we would like to take as a future work.

A threat to construct validity is that large language models are inherently random and thus the case study is not necessarily reproducible. To minimize the threat, we ran ChatGPT five times for each prompt and computed the average performance. We also requested ChatGPT to run multiple times before it presented the final answers.

A special threat to validity is that some of the tested abbreviations/identifiers may have been taken as training data for ChatGPT. As a result, the performance could have been bloated because of the data contamination problem. However, to the best of our knowledge, ChatGPT has not been specially trained for the task of abbreviation expansion. Consequently, even if the identifiers are known to ChatGPT, it is likely that their corresponding expansions have not yet been explicitly told to ChatGPT.

### B. Implications

*1) Easy-to-access generic LLMs have great potential in abbreviation expansion:* Although the state-of-the-art approaches specially designed for abbreviation expansion are as accurate as our approach is, they are not widely employed because of the following drawbacks. First, they are presented as standalone tools and thus they cannot be used unless developers intentionally download and execute the tools. However, few developers know such tools and even fewer would like to install such a special tool for abbreviation expansion. Second, existing tools request expensive code parsing and code analysis, making them less efficient whereas developers are often reluctant to wait. Requesting code parsing also makes such an approach inapplicable to incomplete or illegal source code. In contrast, generic large language models like ChatGPT may serve as the single access point to personal programming assistants, and thus all developers can assess it at any time from anywhere. This advantage makes our approach preferable to the state-of-the-art approaches.

*2) Applying generic LLMs to resolve software engineering tasks like abbreviation expansion often requests essential customization that can only be done by experts in the field:* As suggested by the case study, applying ChatGPT to expand abbreviation without essential customization resulted in poor performance that is significantly lower than the state of the art. To improve the performance, we should analyze the failed cases, figure out the root causes for the failures, and design special measurements by exploiting domain knowledge. Our empirical study also suggests that post-condition checking could be as useful as customized prompt engineering.

## VI. Conclusions and Future Work

Abbreviations are common in source code, and automated expansion of abbreviations has been a long-standing issue. With the significant advances of LLMs in resolving software engineering tasks, LLMs have the potential to expand abbreviations accurately. To validate this assumption, in this paper, we designed and conducted a case study that not only evaluated the potential of ChatGPT in abbreviation expansion, but also revealed the root causes for failure, and proposed a sequence of measures to substantially improve the performance of ChatGPT in abbreviation expansion. As a result, the final approach is as accurate as the state-of-the-art tools specially designed for this task. The special advantage of this approach, i.e., its avoidance of costly source code parsing and analysis, makes it more attractive than the state-of-the-art approaches.

In the future, we would like to repeat the study on larger benchmarks, involving more large language models and additional programming languages. Is also potentially fruitful to apply the same search pattern to other software engineering tasks, e.g., code complete and API recommendation.